\documentclass[12pt]{iopart}

\usepackage{iopams}

\usepackage{graphicx}
\usepackage{subfigure}
\usepackage{dcolumn}
\usepackage{bm}
\usepackage{color}

\newcommand{\delx}{\ensuremath{\Delta x}}
\newcommand{\dely}{\ensuremath{\Delta y}}
\newcommand{\delz}{\ensuremath{\Delta z}}

\newcommand{\sgndelz}{\ensuremath{\mbox{sgn}(\Delta} z)}
\newcommand{\sgndelzhat}{\ensuremath{\mbox{sgn}(\hat{\Delta}} z)}
\newcommand{\hatz}{\ensuremath{\hat{z}}}
\newcommand{\hatdelz}{\ensuremath{\hat{\Delta} z}}
\newcommand{\mhatdelz}{\ensuremath{|\hat{\Delta} z}|}

\def\mhatdelzz#1{\ensuremath{|\hat{\Delta} z(#1)}|}

\newcommand{\hatR}{\ensuremath{\hat{R}}}

\begin{document}

\title{Interaction of a magnetic dipole with a slowly moving electrically conducting plate}

    \author{Evgeny V. Votyakov}
    \address{Computational Science Laboratory UCY-CompSci, \\Department of
    Mechanical and Manufacturing Engineering, \\University of Cyprus, 75
    Kallipoleos, Nicosia 1678, Cyprus}
    \ead{karaul@gmail.com}

    \author{Andr\'{e} Thess}
    \address{Institute of Thermodynamics and Fluid Mechanics,\\
    Ilmenau University of Technology, \\P.~O.~Box 100565, 98684 Ilmenau,
    Germany}
    \ead{thess@tu-ilmenau.de}

\begin{abstract}
We report an analytical investigation of the force and torque
acting upon a magnetic dipole placed in the vicinity of a moving
electrically conducting nonmagnetic plate. This problem is
relevant to contactless electromagnetic flow measurement in
metallurgy and extends previous theoretical works (Thess et al.
\textit{Phys. Rev. Lett.}, \textbf{96}(2006), 164501; \textit{New
J. Phys.} \textbf{9}(2007), 299) to the case where the orientation
of the magnetic dipole relative to the plate is arbitrary. It is
demonstrated that for the case of low magnetic Reynolds number the
three-dimensional distributions of the induced electric potential,
of the eddy currents and of the induced magnetic field can be
rigorously derived. It is also shown that all components of the
force and torque can be computed without any further
approximation. The results of the present work serve as a
benchmark problem that can be used to verify numerical simulations
of more complex magnetic field distributions.

\end{abstract}


\maketitle

\section{Introduction} \label{sec:intro}
When an electrically conductive macroscopic body moves in the
nonuniform magnetic field $\mathbf{B}$ created by an external
source, for instance a permanent magnet, then eddy currents
$\mathbf{j}$ are induced in the body. These eddy currents create
the Lorentz force $\mathbf{F=j\times B}$. The force is directed in
the direction opposite to the direction of the body's translation.
Moreover, the eddy currents induce an additional magnetic field
$\mathbf{b}$ which interacts with the permanent in such a way as
to create a force and a torque acting upon the magnet. Thus, by
measuring these forces and torques one can determine the
parameters of the movement such as the velocity or the direction.
This principle is embodied in a contactless electromagnetic flow
measurement technique called Lorentz force velocimetry
\cite{Shercliff:book:1962}, \cite{Thess:Votyakov:Kolesnikov:2006},
\cite{Thess:Votyakov:NJP:2007}, \cite{Bucenieks:Pamir:2002},
\cite{Bucenieks:Dresden:2005}, \cite{Priede:JApplPhys:2011}, which
permits flow measurement in hot and aggressive fluids like liquid
aluminium or molten steel. When developing measurement systems
embodying Lorentz force velocimetry, so-called Lorentz force
flowmeters, it is necessary to predict the force and torque acting
upon a complex-shaped magnet system by a turbulent liquid metal
flowing in pipes, ducts and open channels. Such predictions are
usually performed by numerically solving the full set of
three-dimensional equations of magnetohydrodynamics
\cite{Davidson:book:2001}. In order to be able to assess the
reliability of such simulations it is necessary to have simple
models that are amenable to rigorous analytic treatment. The goal
of the present paper is to formulate and solve such a model.

The model to be studied in the present work is a generalization of
a previously studied problem \cite{Thess:Votyakov:NJP:2007} to the
case when the orientation of a magnetic dipole is arbitrary. More
precisely, the authors of \cite{Thess:Votyakov:NJP:2007}
investigated the interaction between a moving plate and a magnetic
dipole whose orientation is perpendicular to the surface of the
plate. Here we relax the assumption about the orientation of the
dipole and allow the dipole to be arbitrarily oriented. As will be
shown in Section \ref{sec:problem}, this problem can still be
solved exactly and all components of the torque and the force can
be expressed analytically along with the electric and magnetic
field. We illustrate our results in Section \ref{sec:results}
using some representative plots of the three-dimensional structure
of the eddy currents and the induced magnetic fields. In Section
\ref{sec:conclusion} we summarize our conclusions and indicate a
possible illustrative application of the theory that would be
interesting to investigate.

\section{The problem and its exact solution} \label{sec:problem}

\begin{figure}
  \begin{center}
    \includegraphics[width=0.6\textwidth]{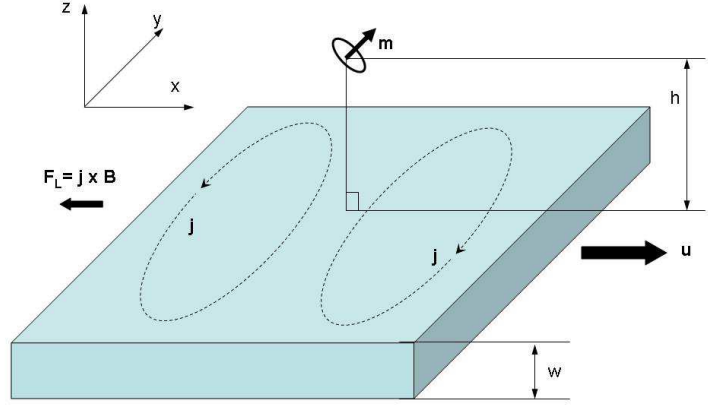}
  \end{center}
  \caption{Sketch of the problem:
  A plate with thickness $w$ moves with constant velocity $u$
  and interacts with an arbitrarily oriented magnetic dipole.}
  \label{Fig:sketch2}
\end{figure}

Consider a single magnetic dipole with dipole moment $\mathbf{m}$
located at a distance $h$ above a fluid or solid layer with
electrical conductivity $\sigma$ and thickness $w$ as shown in
Fig.~\ref{Fig:sketch2}. The dipole is assumed to be arbitrary
oriented in space. For brevity, if  it is not specified otherwise,
we assume that the index $n$  denotes one of the Cartesian axes,
$\mathbf{m}=m\mathbf{e}_n$. The general case can be considered as
a linear combination of the three generic cases, $n=x,y,z$, any
general orientation $a$ can be expressed as
$a=\bm{m}\cdot\{\bm{a}^{(n)}\}^T$, where
$\{\bm{a}^{(n)}\}=(a^{(x)},a^{(y)},a^{(z)})$ is the tensor
containing components for each $n=x,y,z$ dipole orientation. The
layer moves with an uniform  velocity $\textbf{u}=u\textbf{e}_x$
and extends in the $z$ direction from $z_d$ to $z_u$. The
thickness of the layer is then given by  $w=|z_u-z_d|$. Here, the
indices "$d$ " and "$u$ " stand for "downward side" and "upward
side", respectively. The borders of the layer in the $x$ and $y$
direction are supposed to be at infinity.  This corresponds to the
assumption that the distance $h$ is much less than the horizontal
size of the moving solid layer. We further assume that plate or
the fluid moves slowly which is equivalent to the assumption that
magnetic Reynolds number $Rm=\mu_0\sigma uw$ is small. In
magnetohydrodynamics (see e.g. \cite{Roberts:1967},
\cite{Davidson:book:2001}) this case is referred to as the
quasistatic approximation and implies that the magnetic field
associated with the eddy currents is much smaller than the applied
magnetic field. The assumption $Rm\ll 1$ makes the present problem
amenable to rigorous analytic treatment. For instance, the
magnetic Reynolds number of liquid aluminium ($\sigma\approx
3\times 10^6$) flowing with a velocity $u=1$ m/s in a pipe with a
diameter of 0.1 m is approximately 0.3. In what follows we will
use mainly Cartesian coordinates and will not apply an integration
of Bessel functions in cylindrical coordinates as we did before in
\cite{Thess:Votyakov:NJP:2007}. Our goal is to derive explicit
compact algebraic expressions, which are useful for fast analytic
calculations, for the primary magnetic field $\mathbf{B(r)}$, the
electric potential $\phi(\mathbf{r})$, the eddy currents
$\mathbf{j(r)}$ and the secondary magnetic field $\mathbf{b(r)}$.
Then, all other quantities such as forces and torques acting upon
the dipole can be readily  written down.

\subsection{Primary magnetic field} \label{sec:primary}

The first step in the solution of our problem is the specification
of the magnetic field of the dipole which is referred to as the
primary magnetic field. Let the magnetic dipole be at the point
$\mathbf{r}_0=(x_0,y_0,z_0)$ and we wish to calculate the magnetic
field $\mathbf{B(r)}$ at a point $\mathbf{r}=(x,y,z)$. Before
presenting formulae for the primary magnetic field, we introduce
the following short notations of the vector
$\mathbf{R}=\mathbf{r}-\mathbf{r}_0$:
\begin{eqnarray}\label{eq:MF:identities}
    \delx &=& x-x_0, \hskip 1cm \dely = y-y_0, \hskip 0.7cm \delz = z-z_0, \nonumber\\
    \Delta r^2 &=& \delx^2 + \dely^2, \hskip 0.9cm
    R^2=\Delta r^2+\delz^2, \nonumber\\
    \frac{1}{R} &=& \frac{\partial}{\partial z} \left[\tanh^{-1}\frac{\delz}{R}\right],
    \hskip 0.7cm \nabla\left[\frac{1}{R}\right]=-\frac{\mathbf{R}}{R^3}
\end{eqnarray}
Now, the magnetic field of a magnetic dipole can be represented as
\cite{Jackson:Book:1999}:
\begin{eqnarray}\label{eq:MF:dipole}
    \mathbf{B(r)}
    &=& \frac{\mu_0}{4\pi}\nabla\times\left[{\bf m}\times\frac{{\bf R}}{R^3}\right]
    =  \frac{\mu_0}{4\pi} \nabla \times\left[ - {\bf m} \times \left(
    \nabla\frac{1}{R}\right) \right] \nonumber \\
    &=& \frac{\mu_0}{4\pi} \nabla\times\left[
    \nabla\times\left( \frac{{\bf m}}{R}\right) -
    \frac{1}{R}\nabla\times{\bf m}
    \right]
    = \frac{\mu_0}{4\pi} \nabla\left[ {\bf m}\cdot\nabla\left(\frac{1}{R}\right)\right],
\end{eqnarray} These expressions appear more complicated than the formula normally used.
However, the present notation will turn out to be convenient for
the further analysis and it is the key to a compact representation
of all fields. Then it follows, by taking (\ref{eq:MF:identities})
and $\mathbf{m}=m\mathbf{e}_n$ into account, that  the components
of the primary magnetic field can be conveniently given as:
\begin{eqnarray}\label{eq:MF:components}
    B_k = \mu\,\partial_k \partial_n \left[\frac{1}{R}\right]
    =\mu\,\partial_k\partial_n\partial_z  \left[\tanh^{-1}\frac{\Delta
    z}{R}\right], \hskip 0.5cm k=x,y,z,
\end{eqnarray} where we introduced the abbreviation
$\mu=m\,\mu_0/4\pi$, and $\partial_n(.)$ is the partial derivative
with respect to  the coordinate specified by the direction
$\bm{n}$ of the dipole. If the dipole $\mathbf{m}$ is arbitrary
oriented in space,
$\mathbf{m}=m_x\mathbf{e}_x+m_y\mathbf{e}_y+m_z\mathbf{e}_z$, then
$\partial_n \equiv \mathbf{m}\cdot\nabla$. For instance, for  the
dipole oriented parallel to the  $z$-axis, $\bm{n}=\bm{e}_z$ and
$n=z$, and the components of the primary magnetic field are:
\begin{eqnarray}\label{eq:MF:componentsmz}
    B_x = 3\mu\,\frac{\Delta x\,\Delta z}{R^5}, \hskip 0.5cm
    B_y = 3\mu\,\frac{\Delta y\,\Delta z}{R^5}, \hskip 0.5cm
    B_z = \mu\left[\frac{3\,\Delta z^2}{R^5}-\frac{1}{R^3}\right].
\end{eqnarray}

\subsection{Electric potential} \label{sec:potential}

The derivation of the electric potential is an extension of the
method developed earlier in \cite{Thess:Votyakov:NJP:2007}, and it
is presented also in \cite{Priede:JApplPhys:2011}.

The electric potential $\phi$ and eddy currents $\mathbf{j}$ are
governed by Ohm's law for an electrically conducting material
moving with velocity $\mathbf{u}$,  which can be written as:
\begin{equation}\label{eq:Ohm}
  \mathbf{j} =\sigma(-\nabla\phi + \mathbf{u} \times \mathbf{B}),
  \hskip 0.5cm \nabla\cdot\mathbf{j}=0.
\end{equation} (The second expression is not Ohm's law and does
imposes the requirement that there is no source or sink of
electric currents in the moving plate.) In general, the magnetic
field in (\ref{eq:Ohm}) is the sum of the primary magnetic field
and the  magnetic field associated with the eddy currents,
referred to as the secondary magnetic field. As explained earlier,
we assume that the magnetic Reynolds number $Rm=\mu_0\sigma vh$,
which characterizes the ratio between the secondary and primary
field, is small, so we can approximate the magnetic field in Ohm's
law by the primary field given by (\ref{eq:MF:components}).

The primary magnetic field is solenoidal,
$\nabla\times\mathbf{B}=0$, and the velocity $\mathbf{u}$ is
constant. By applying the divergence operator to Ohm's law
(\ref{eq:Ohm}) we therefore  obtain the following equation for the
electric potential:
\begin{equation}\label{eq:laplace}
  \Delta \phi =\nabla \cdot\left[\mathbf{u} \times \mathbf{B}\right]=
   \mathbf{B} \cdot \left[\nabla \times \mathbf{u}\right]-
   \mathbf{u} \cdot \left[\nabla \times \mathbf{B}\right]=0.
\end{equation}Hence, $\phi$ is a harmonic function obeying Laplace's equation, $\Delta \phi =0$.
Any harmonic function in three-dimensional space can be expressed
through derivatives or integrals of the function $1/R$ defined in
(\ref{eq:MF:identities}), where the specific expression for $\phi$
depends on boundary conditions. In our case, the domain is
infinitively large in  $x$ and $y$ directions, and finite in the
$z$ direction. At the upper ($z=z_u$) and lower ($z=z_d$) surfaces
of the plate, the vertical electric current must be zero,  which
is expressed as $j_z|_{z=z_s}=0$, and hence,
$\sigma(-\partial_z\phi+uB_y)|_{z=z_s}=0$, where the index s
stands for either $s=u$ or $s=d$. Hence, we can directly express
$\partial_z\phi$ by setting $j_z=0$ everywhere in the layer:
\begin{equation}\label{eq:BC}
  \partial_z\phi = u\,B_y=
  u\mu\,\partial_y\partial_n\partial_z \left[\tanh^{-1}\frac{\Delta z}{R}\right].
\end{equation}where the last equality directly follows from
(\ref{eq:MF:components}). Now, if we change the order of
differentiation and put $\partial_z$ as the first derivative, then
the integration is trivial, and we come immediately to the final
result:
\begin{eqnarray}\label{eq:potential}
  \phi(x,y,z) &=& u\mu\,\left\{\partial_y\partial_n \left[\tanh^{-1}\frac{\Delta
  z}{R}\right]+\phi_\infty(x,y)\right\},
\end{eqnarray} where the second term $\phi_\infty(x,y)$ appears
as constant of the integration and so it does not depend on $z$.
This term is determined from the additional boundary condition
specifying that the electric potential vanishes when the dipole is
at infinite distance from the layer. The result is:
\begin{eqnarray}\label{eq:potentialinf}
  \phi_\infty(x,y) &=& -\lim_{z_{0}\rightarrow\infty}
        \partial_y\partial_n
        \left[\tanh^{-1}\frac{\delz}{R}\right]
        \nonumber \\
        &=&-\partial_n\left\{\lim_{z_{0}\rightarrow\infty}
        \partial_y \left[\tanh^{-1}\frac{\delz}{R}\right]\right\}
    \nonumber \\
    &=& \partial_n\left\{\sgndelz\frac{\dely}{r}\right\}
    = \sgndelz \,\partial_y\partial_n  \ln r,
\end{eqnarray} where $\mbox{sgn}(z)=z/|z|$ is the sign function  with the property that
$\mbox{sgn}(\Delta z)=1$ for $z>z_0$ and $\mbox{sgn}(\Delta z)=-1$
for $z<z_0$. One can readily verify that $\phi$ given by
(\ref{eq:potential}) is the solution of (\ref{eq:laplace}). To do
this, we notice that $\phi=\phi(1/R)$, where $1/R$ is given by
(\ref{eq:MF:identities}) and $\Delta[1/R]=\Delta[\tanh^{-1}(\Delta
z/R)]=0$, hence, $\Delta \phi=0$ too.

In \ref{app:stream} it is shown that the  components of electric
current can be expressed through a stream function $\psi$ defined
as $\mathbf{j}=\nabla\times(\psi\mathbf{e}_z)$. This stream
function is given by:
\begin{eqnarray}\label{eq:psi:intro}
  \psi(x,y,z)&=& -u\mu\sigma\left\{
    \partial_x \partial_n \left[\tanh^{-1}\frac{\Delta
    z}{R}\right]+\psi_\infty(x,y)\right\},\\
    \psi_\infty(x,y) &=& -\lim_{z_{0}\rightarrow\infty}
   \partial_x\partial_n \left[\tanh^{-1}\frac{\Delta
  z}{R}\right] =\sgndelz \,\partial_x\partial_n \ln r.\nonumber
\end{eqnarray} In the short notation both $\phi$ and $\psi$ can be
conveniently represented by means of the differentiation of the
same auxiliary function $\Phi$ as:
\begin{eqnarray}\label{eq:phi_psi_short}
    \phi^{(n)}=u\mu\,\partial_y\partial_n \Phi, \hskip 1.5cm
    \psi^{(n)}=-u\mu\sigma\,\partial_x\partial_n \Phi,
    \\
    \Phi=\sgndelz\left\{\tanh^{-1}\frac{|\delz|}{R}+\ln r\right\} \nonumber,
\end{eqnarray}where an upper index $n$ specifies explicitly the
direction of the magnetic dipole.

Because the electric currents are two-dimensional, the solution
(\ref{eq:phi_psi_short}) describes an electric potential and
current stream function for any horizontal plane of the moving
layer. Explicit formulae for $\psi$ and $\mathbf{j}$ are given in
\ref{app:stream}. These formulae can be immediately used for
calculation without performing cumbersome differentiation.

\subsection{Secondary magnetic field and scalar potential}

The eddy currents $\mathbf{j}$ of the moving plate induce a
 magnetic field $\mathbf{b}$  which we refer to as the secondary magnetic field.
When $\mathbf{j}$ is time-independent as in our case, the
secondary magnetic field $\mathbf{b}(x,y,z)$ at a point
$\mathbf{r}=(x,y,z)$ is uniquely determined by Ampere's law
\begin{equation}\label{eq:Ampere}
  \mathbf{j}(x,y,z)=\frac{1}{\mu_0}\nabla\times\mathbf{b}(x,y,z).
\end{equation} On the other hand, $\mathbf{j}$ can be determined with the aid of
the stream function which has been already above calculated,
Eq.(\ref{eq:phi_psi_short}):
\begin{eqnarray*}
    \mathbf{j}(x,y,z)=\Theta(z)\nabla\times\left[\mathbf{e}_z\psi(x,y,z)\right].
\end{eqnarray*}Here the auxiliary step function $\Theta(z)$ is
introduced  in order to ensure that  eddy currents outside of the
plate are zero. For instance, in the case of the layer of finite
thickness, the function $\Theta(z)$ is defined as:
\begin{eqnarray*}
    \Theta(z) =\left\{ \begin{array}{ll}
    1& \mbox{for $z_d \leq z \leq z_u$};
    \\
    0& \mbox{for $z < z_d$ and $z_u < z$},\end{array} \right.
\end{eqnarray*}where $z_d$ and $z_u$ are the lower and upper borders of the
plate.  In the general case, the secondary field can be
represented completely as a sum of the stream function and a
gradient of a scalar function $a$ in the form:
\begin{equation}\label{eq:Ampere:gradient}
  \mathbf{b}(x,y,z)=\Theta(z) \, \mu_0\,\psi(x,y,z)\,\mathbf{e}_z + \mu_0\nabla a(x,y,z),
\end{equation}

Function $\Theta(z)$ enforces the condition that outside the plate
the secondary field $\mathbf{b}$ is solely a gradient of the
scalar potential $a$. This automatically ensures that
$\nabla\times\mathbf{b}=0$, and then the problem how to calculate
$\mathbf{b}$ is reduced into the problem how to calculate the
scalar potential $a$. In order to accomplish this task one has to
employ the constraint that $\mathbf{b}$ is divergence-free. By
taking the divergence of (\ref{eq:Ampere:gradient}), the
divergence-free property, $\nabla\cdot\mathbf{b}=0$, results in
the following equation:
\begin{eqnarray}\label{eq:Ampere:laplace}
    \Delta a=-\Theta(z) \, \partial_z\psi
    = \Theta(z)\,u\mu\sigma\,\partial_x \partial_n\left[\frac{1}{R}\right],
\end{eqnarray}where the second equality directly follows from
Eq.~(\ref{eq:phi_psi_short}). So one has to solve now  equation
(\ref{eq:Ampere:laplace}) with the boundary conditions $|\nabla
a|\rightarrow 0$ which describes that the secondary magnetic field
vanishes at infinity.

It is possible to solve Eq.~(\ref{eq:Ampere:laplace}) in a
straightforward way for any specific $n=x,y,z$. This requires
explicit partial derivatives with respect to $x$ and $n$ and then
cumbersome symbolic calculations for each separate index $n$ by
recruiting series expansion with Bessel functions and further
integration. Such procedure  has been performed in
\cite{Thess:Votyakov:NJP:2007} for the particular case $n=z$, see
also \cite{Priede:JApplPhys:2011}. However, there exists another
technique to provide a general formula for arbitrary general $n$.
For this we notice that a derivative of any function $f(R)$, which
depends on $R$ solely, with respect to $n=x,y,z$ can be taken as a
derivative, with the opposite sign, with respect to
$n_0=x_{0},y_{0},z_{0}$, correspondingly, $\partial_n
f(R)=-\partial_{n_{0}}f(R)$, where $R$ is defined in
(\ref{eq:MF:identities}). In order to use this property of $R$ we
represent the scalar potential $a$ by means of a generating (or
primitive) function $A$ twice differentiated with respect to the
coordinates of the dipole,
$a=u\mu\sigma\,\partial_{x_{0}}\partial_{n_{0}} A$. We insert the
above formula for $a$ into (\ref{eq:Ampere:laplace}) and obtain
the simpler Poisson's equation independent of $n$ to determine $A$
as:
\begin{equation}\label{eq:Ampere:Poisson:A}
    \Delta A = \Theta(z)\frac{1}{R}.
\end{equation} This equation can be readily solved without cumbersome integration and
series expansion as we shall demonstrate below. Once we obtain the
primitive function $A$, the secondary magnetic field can be
represented by means of triple differentiation  in the form:
\begin{equation}\label{eq:Ampere:b:A}
    \mathbf{b}^{(n)}=\mu_0\mu\sigma
    u \nabla\left[\partial_{x_{0}}\partial_{n_{0}}A\right].
\end{equation}

\subsection{Scalar potential and secondary magnetic field for an infinitely thin layer}

Because the problem under consideration is linear, we treat the
case of an infinitely thin layer separately. Once a solution for
this case is found, all other cases can be presented through a
superposition of thin layers.

Let  $\hat{z}$ be a location inside a thin moving layer with
electric currents induced by an external magnetic dipole. Let us
assume that the thickness of the layer is  small compared to the
distance between the dipole and the layer  which is expressed by
the condition $w=|z_u-z_d|\ll h$. Finally let us introduce
$\hat{z}=|z_u+z_d|/2$. In this case, the function $\Theta(z)$,
which is needed  to set the position of the layer, must be taken
in the limit $z_d\rightarrow z_u$. It gives
$\Theta(z)=w\,\delta(z-\hat{z})$, where $w$ is the thickness of
the layer which is inserted for the proper physical
dimensionality. (Function $\Theta(z)$ is dimensional in the case
of the infinitely thin layer, but $w$ will be omitted for other
cases because there the proper dimensionality follows  directly
from an integration over the plate thickness.)

Let $A_0$ be the primitive function for the thin layer. Here the
index 0 is inserted to distinguish the present case from other
cases. As follows from Eq.(\ref{eq:Ampere:Poisson:A}), everywhere
in the space $\Delta A_0=0$ except for $z=\hat{z}$. This means
that $A_0$ is a continuous harmonic function showing a
discontinuity in the first derivative with respect to  $z$ at
$z=\hat{z}$. As shown in \ref{app:proofA0}, the solution of
(\ref{eq:Ampere:Poisson:A}) is:
\begin{eqnarray}\label{eq:A:thinlayer}
  A_0(\hat{z}) &=& \frac{w}{2}\left\{
        \tanh^{-1}\frac{\mhatdelzz{\hatz}}{\hat{R}(\hatz)}+\ln r\right\},\\
   \hat{R}(\hatz)^2&=&r^2+\mhatdelzz{\hatz}^2,
  \hskip 0.3cm \mhatdelzz{\hatz}=|\hatz-z_0|+|\hatz-z|. \nonumber
\end{eqnarray} Notation $\mhatdelzz{\hatz}$ looks ugly but
this is because we want to stress an analogy in between functions
$R$ and $\hat{R}$ as explained below. In the given case of a
infinitely thin layer, it is superfluous to put argument $\hatz$
in the round brackets of the function $\hat{\Delta} z$, but this
argument will be necessary afterwards, when we shall extend these
results  to the case of a layer with finite thickness.

Note, that  $A_0$ in (\ref{eq:A:thinlayer}) and $\Phi$ in
(\ref{eq:phi_psi_short}) are almost identical, that is,
$(w/2)\Phi$ corresponds to $\sgndelzhat A_0$, with the mapping
$R\leftrightarrow\hat{R}$ and $\Delta z\leftrightarrow\hat{\Delta}
z$. Moreover, further analysis shows that this correspondence is
valid not only for the entire functions $A_0$ and $\Phi$, but also
for all their derivatives with respect to $n=x,y,z$ and
$n_0=x_0,y_0,z_0$, that is, $(w/2)\partial_n\Phi \leftrightarrow
\sgndelzhat\partial_n A_0$,
$\partial_{n}\Phi=-\partial_{n_{0}}\Phi$, and
$\partial_{n}A_0=-\partial_{n_{0}}A_0$. The only exception is
$\partial_{z}A_0=\partial_{z_{0}}A_0$ because $(z-z_0)$ is taken
in $A_0$ as an absolute value. By taking this property into
account and recalling that both the eddy currents $\mathbf{j}$ and
secondary magnetic field $\mathbf{b}$ can be represented through a
triple differentiation as:
\begin{eqnarray*}
    \frac{\bm{j}}{\mu\sigma u}=
    -\nabla\times\left[\partial_x\partial_n \Phi \right]\bm{e}_z, \hskip
    0.5cm
    \frac{\bm{b}}{\mu_0\mu\sigma u}=
    \nabla\left[\partial_{x_{0}}\partial_{n_{0}} A_0 \right].
\end{eqnarray*} Therefore without computing derivatives, it is possible to build up the following
correspondence between the secondary magnetic field and eddy
currents:
\begin{eqnarray}
    b^{(n)}_{x} \leftrightarrow  \frac{\sgndelzhat}{2} j^{(n)}_y , \hskip 0.5cm
    b^{(n)}_{y} \leftrightarrow -\frac{\sgndelzhat}{2} j^{(n)}_x \hskip 0.25cm
    \mbox{  for  } n=x,y \label{eq:bxy:thin:layer1}
     \\
    b^{(z)}_{x} \leftrightarrow -\frac{\sgndelzhat}{2} j^{(z)}_y, \hskip 0.5cm
    b^{(z)}_{y} \leftrightarrow  \frac{\sgndelzhat}{2} j^{(z)}_x \hskip 0.25cm
    \mbox{  for  } n=z. \label{eq:bxy:thin:layer2}
\end{eqnarray}(The difference in sign for $n=z$ is because $\partial_z\Phi=-\partial_{z_{0}}\Phi$ while $\partial_z
A_0=\partial_{z_{0}} A_0$.) The remaining $z$-components of the
secondary field are:
\begin{eqnarray}
    b^{(x)}_{z} \leftrightarrow \frac{\sgndelzhat}{2} j^{(z)}_y,\hskip 0.3cm
    b^{(y)}_{z} \leftrightarrow \frac{\sgndelzhat}{2} j^{(z)}_x,\hskip
    0.3cm \nonumber \\
    b^{(z)}_{z} \leftrightarrow \frac{\sgndelzhat}{2}\left\{j^{(x)}_y-j^{(y)}_x\right\}.
    \label{eq:bxy:thin:layer3}
\end{eqnarray}The last relationship for $b^{(z)}_{z}$ is the
conjecture of the fact that the $\Phi$ is a harmonic function,
i.e. $\Delta \Phi =0$, hence, $(j^{(x)}_y-j^{(y)}_x)/(\mu\sigma u)
=\partial_x\left(\partial_x\partial_x+\partial_y\partial_y\right)\Phi
=\left[-\partial_x\partial_z\partial_z\Phi\right]$ and
$(w/2)\,\partial_x\partial_z\partial_z\Phi \leftrightarrow
\sgndelzhat\,\partial_x\partial_z\partial_z A_0$. According to the
mapping (\ref{eq:bxy:thin:layer1}-\ref{eq:bxy:thin:layer3}), the
tensor $\{\bm{b}^{(n)}\}$ of the secondary magnetic field can be
now written down by means of the tensor $\{\bm{j}^{(n)}\}$ given
in (\ref{eq:j}) for eddy currents. This tensor is presented in
\ref{app:tensor:induced:thin} and can be straightforwardly used in
algebraic calculations in contrast to the implicit results
published by Priede \cite{Priede:JApplPhys:2011}.

\subsection{Secondary magnetic field for half-space and for the layer with finite thickness}

Because the problem under consideration is linear, the secondary
field of any moving plate can be represented by summing all the
fields originating from the layers of the plate derived in the
preceding section. Mathematically, this means an integration of
the secondary magnetic field tensor, cf.
Eq.~(\ref{eq:secondaryfield}), over the thickness of the plate.
The limits of the integration in the case of half-space are $z_u$
and $z_d\rightarrow -\infty$. Then, the integration should be
carried out for the $P_k$ coefficients only because the functions
$f_k$ and $g_k$ do not depend on $z$. By performing this procedure
formally, we obtain
\begin{eqnarray}\label{eq:Pk:thickplate0}
    P_{\infty,k}=\int^{z_u}_{-\infty}P_{0,k}(\hat{z})\,\,d\hat{z},
\end{eqnarray} where $P_{0,k}$ and $P_{\infty,k}$ are tensor coefficients of an infinitely
thin layer and half-space, correspondingly. Explicit formulae for
$P_{\infty,k}$ are presented in \ref{app:tensor:induced:thick}.

When the plate is of finite thickness, $z_d \le \hat{z}\le z_u$,
the coefficients $P_{k}$ are obviously expressed through the
coefficients for half-space in the form:
\begin{eqnarray}\label{eq:Pk:finiteplate}
    P_k(z_u,z_d)&=&\int^{z_{u}}_{z_d}
    P_{0,k}({\hat{z}})\,\,d\hat{z} = \int^{z_{u}}_{-\infty} P_{0,k}({\hat{z}})\,\,d\hat{z}
    -\int^{z_{d}}_{-\infty} P_{0,k}({\hat{z}})\,\,d\hat{z} \nonumber
    \\
    &=& P_{\infty,k}(z_u) - P_{\infty,k}(z_d),
\end{eqnarray} where $P_{\infty,k}(z_u)$ and
$P_{\infty,k}(z_d)$ are computed according to
(\ref{eq:Pk:thickplate}) with $\hat{\Delta}z$ and $\hat{R}$ taken
at points $z_u$ and $z_d$.

\subsection{Magnetic potential energy, force and torque}
In the general case, the potential energy $U$ of a magnetic dipole
$\bm{m}$ in a magnetic field $\bm{B}$ is $U=-\bm{m}\cdot\bm{B}$,
see e.g. \cite{Griffiths:book:1989}. In our case, the magnetic
field is the induced secondary magnetic field
$\bm{b}=\mu_0\mu\sigma u\left[ \bm{m}\cdot \{\bm{b}^{(n)}\}^T
\cdot \bm{m} \right]$, hence, the potential energy of the dipole
is written down as:
\begin{eqnarray}\label{eq:potentialenergy}
  U &=& -\mu_0\mu\sigma u\left[ \bm{m}\cdot \{\bm{b}^{(n)}\}^T \cdot
      \bm{m} \right]. \nonumber \\
    &=& -\mu_0\mu\sigma u\left[m_x^2\, b^{(x)}_x + m_y^2\, b^{(y)}_y +
    m_z^2\, b^{(z)}_z  +  2\,m_x\,m_y b^{(y)}_x   \right],
\end{eqnarray}where the last relationship is obtained by taking
into account a symmetry of the secondary field tensor
$b^{(x)}_y=b^{(y)}_x$, $b^{(z)}_y=-b^{(y)}_z$, and
$b^{(z)}_x=-b^{(x)}_z$.

Force $\bm{F}$ acting upon the dipole can be calculated as a
negative gradient of the potential energy:
\begin{eqnarray}\label{eq:force0}
  \bm{F}&=& -\nabla U, \hskip 0.3cm F_i=\mu_0\mu\sigma u
    \left[ \bm{m}\cdot\{\bm{f}^{(n)}_i\}^T \cdot \bm{m}\right],
\end{eqnarray}where a force field tensor $\{\bm{f}^{(n)}_i\}$ is a partial derivative
of the secondary magnetic field tensor with respect to the
coordinate axis $i=x,y,z$, i.e.:
\begin{eqnarray}\label{eq:force1}
  \{\bm{f}^{(n)}_i\}=\partial_i \{\bm{b}^{(n)}\}.
\end{eqnarray}Because the tensor $\{\bm{b}^{(n)}\}$ has
been found above, algebraic formulae for the tensor
$\{\bm{f}^{(n)}_i\}$ can be computed as well. They are presented
by Eq.~(\ref{eq:forcex}-\ref{eq:forcez}). Notice, that these
tensors contain 27 terms (3 times 9), while due to symmetry and
algebraic simplifications they can be given by means of linear
combinations of five coefficients $Q_{k}$, $k=0,\dots,4$.

The torque $\bm{T}$ acting upon the dipole is:
\begin{eqnarray}\label{eq:torque}
  \bm{T} = \bm{m}\times\bm{b}
         = -\mu_0\mu\sigma u\left( \bm{m}\times \left[\{\bm{b}^{(n)}\}^T \cdot
      \bm{m} \right]\right).
\end{eqnarray} It can be computed also also by means of the
tensor $\{\bm{b}^{(n)}\}$ presented explicitly in
(\ref{eq:secondaryfield}).

Above results for the potential energy, force and torque are of
the same structure when the secondary magnetic field is induced by
a discrete ensemble of magnetic dipoles or by a continuous
macroscopic magnet. (The latter can be obtained by an integration
over the space occupied by the magnet.) However in our simplest
case, in order to calculate the force and torque acting upon a
single dipole, we just put $r=0$ and $z=z_0$ in all the tensors
found already. Mathematically, these tensors are sums of the
terms, which have functions $f_k$ and $g_k$ as factors with
coefficients $P_k$ or $Q_k$. The functions $f_k$ and $g_k$ are
equivalent to $\cos k\phi$ and $\sin k\phi$, correspondingly,
here, $\tan\phi=\dely/\delx$, and, hence, $f_k$ and $g_k$ are zero
at $r=0$ for $k\ge 1$. Hence, at $r=0$, all the tensor terms
having factors $P_k$ or $Q_k$ with $k\ge 1$ must vanish due to
their products with $f_k$ and $g_k$. This greatly simplifies all
the formulae, and we obtain finally:
\begin{eqnarray}
  \bm{b} &= \mu_0\mu\sigma u \left.P_0\right|_{r=0,z=z_0}\left\{-m_z\bm{e}_x + m_x\bm{e}_z
  \right\},  \label{eq:secfieldr0}  \\
  \bm{F} &= \mu_0\mu\sigma u \left.Q_0\right|_{r=0,z=z_0}\left\{(3\,m_x^2+m_y^2+4\,m_z^2)\bm{e}_x + 2\,m_x m_y\bm{e}_y
  \right\},
  \label{eq:forcer0}\\
  \bm{T} &= \mu_0\mu\sigma u \left.P_0\right|_{r=0,z=z_0}\left\{m_x m_y \bm{e}_x - (m_x^2+m_z^2)\bm{e}_y + m_y m_z\bm{e}_z
  \right\},
  \label{eq:torquer0}
\end{eqnarray} where
the non-zero tensor coefficients $P_0$ and $Q_0$ at $r=0$ and
$z=z_0$ are presented in \ref{app:P0Q0} in formulae
(\ref{eq:P0:thin}, \ref{eq:Q0:thin}) for a thin layer,
(\ref{eq:P0:thick}, \ref{eq:Q0:thick}) for a half space, and
(\ref{eq:P0:finite}, \ref{eq:Q0:finite}) for the layer with finite
thickness.

\section{Spatial structure of eddy currents and secondary magnetic
field}\label{sec:results}
In order to illustrate the analytical results of previous
sections, we plotted eddy currents and the secondary magnetic
field by using the formulae derived above, Eq.~\ref{eq:psi}) and
(\ref{eq:secondaryfield}). The eddy currents below are shown as
contour lines of the eddy-current stream function $\psi$ given by
Eq.~(\ref{eq:psi}). Solid lines correspond to positive $\psi$ and
indicate counterclockwise direction of the eddy current;  whereas
dotted lines are for negative contour levels and imply clockwise
direction of the eddy current. The secondary magnetic field is
given in the form of selected three-dimensional magnetic field
lines  computed by using Eq.~(\ref{eq:secondaryfield}). According
to the right-hand rule, the secondary magnetic field lines emanate
from the solid positive loops and penetrate the negative dotted
loops.

\begin{figure}[ht]
    \centering
    \subfigure[]{
    \includegraphics[width=0.45\textwidth]{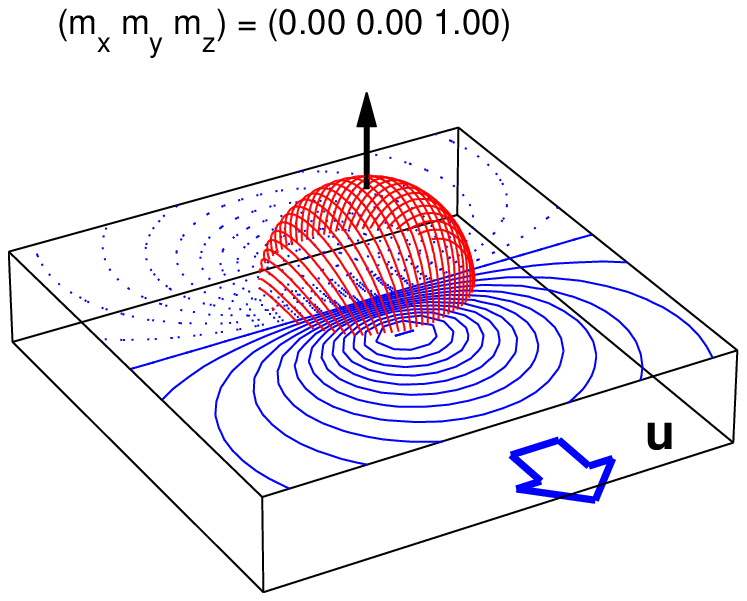}
    \label{fig:th000phi000}
    }
    \subfigure[]{
    \includegraphics[width=0.45\textwidth]{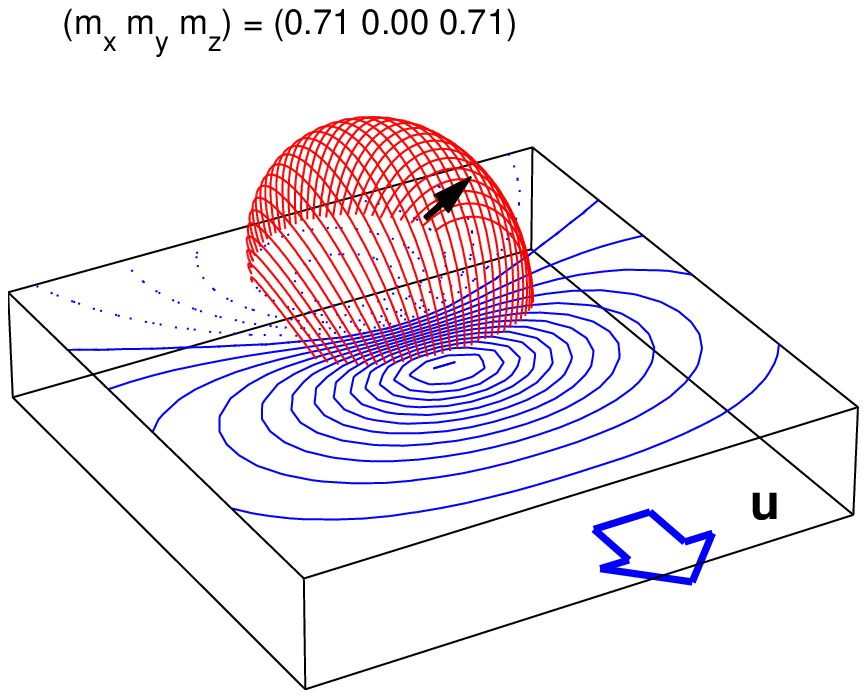}
    \label{fig:th025phi000}
    }\\
    \subfigure[]{
    \includegraphics[width=0.45\textwidth]{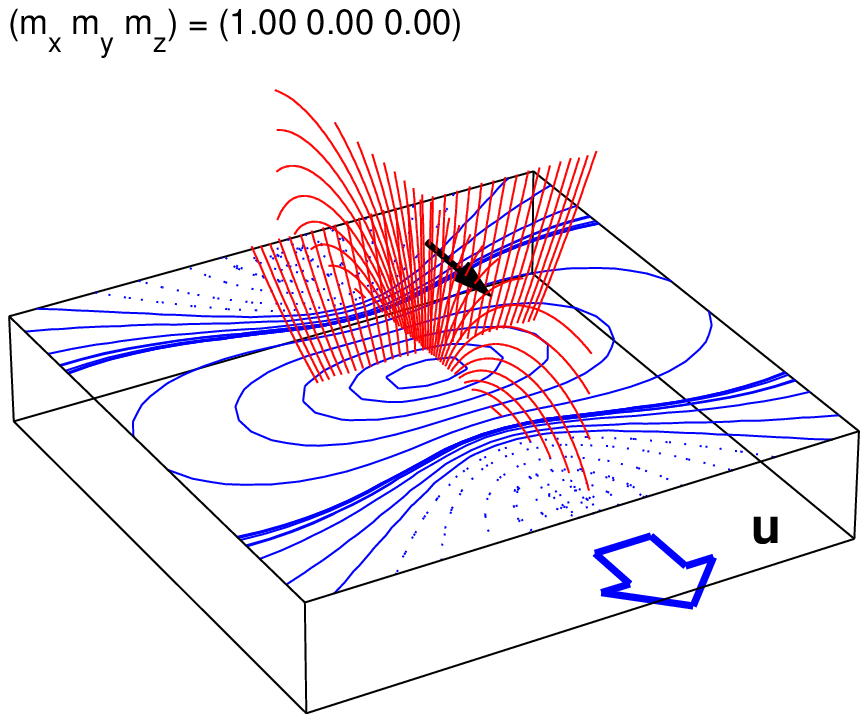}
    \label{fig:th050phi000}
    }
    \hskip 1cm
    \subfigure[]{
    \includegraphics[scale=0.4]{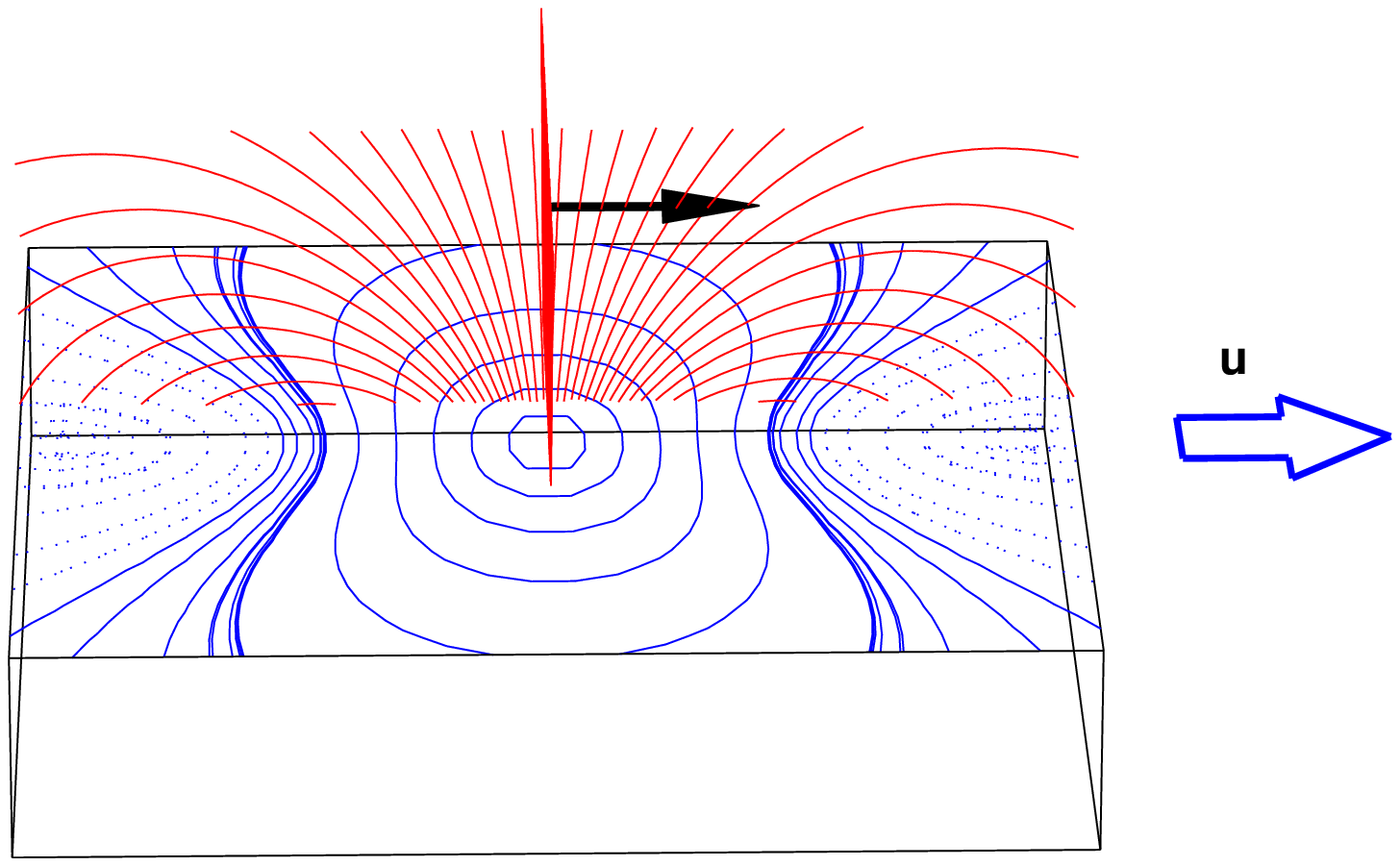}
    \label{fig:th050phi000_3}
    }\\
    \subfigure[]{
    \includegraphics[width=0.45\textwidth]{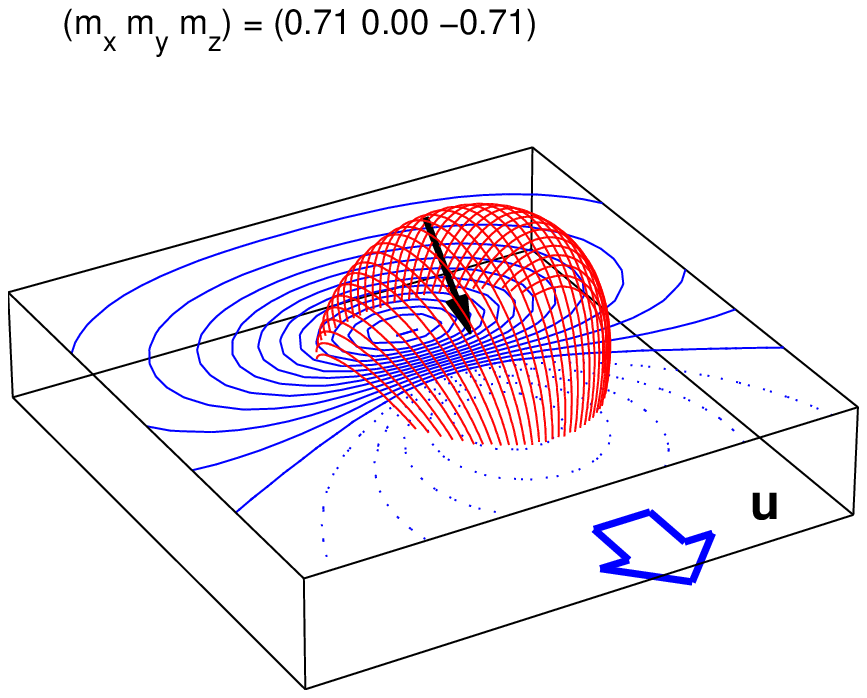}
    \label{fig:th075phi000}
    }
    \subfigure[]{
    \includegraphics[width=0.45\textwidth]{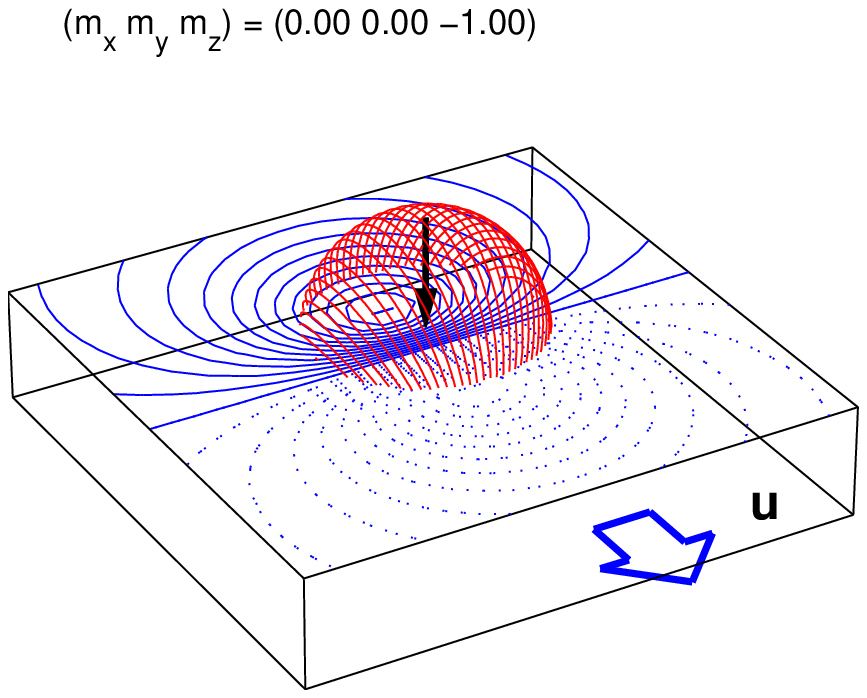}
    \label{fig:th100phi000}
    }
    \caption[Optional caption for list of figures]{
    Eddy currents and secondary magnetic field for the case
    $m_y=0$, i.e. when the magnetic dipole is located in the plane spanned by
    the velocity vector and the direction normal to the surface of the
    plate. The points where the secondary
    magnetic field lines emanate are selected manually inside and nearby the
    eddy-current loops  in order to guide the eye.}
    \label{fig:phi000}
\end{figure}

Figure~\ref{fig:phi000} shows behavior at $m_y=0$ and continuously
changing $m_x$ and $m_z$ provided that $m_x^2+m_y^2+m_z^2=1$. This
corresponds to the case when the dipole is confined to the plane
spanned by the velocity vector and the direction is normal to the
surface of the plate. The results of Fig.~\ref{fig:th000phi000}
have already been given in \cite{Thess:Votyakov:NJP:2007}. In this
case, $m_x=m_y=0, m_z=1$, and there are two loops of eddy currents
whose centers are arranged parallel to the  direction of the plate
movement. The secondary magnetic field lines form a cage. When
$m_x$ contribution increases, both loops deform and the positive
loop grows in its size and shifts under the dipole as shown in
Fig.~\ref{fig:th025phi000}. Then, the case of
Fig.~\ref{fig:th050phi000} corresponds to a magnetic dipole, which
is oriented along the $x$-direction, i.e. $m_x=1$, $m_y=0$,
$m_z=0$, where the positive eddy current loop is located
completely under the dipole, and two negative loops are at the
periphery. As a result, the three-dimensional lines of the
secondary magnetic field are either completely in the $y-z$ plane,
if they start at $x=0$, or the lines are closed at the peripheral
negative loops if they start at $y=0$. This behavior is emphasized
by Fig.~\ref{fig:th050phi000_3} which refers to the same
parameters as Fig.~\ref{fig:th050phi000} but plotted from a
different perspective. Then, Fig.~\ref{fig:th075phi000} is a
mirror of Fig.~\ref{fig:th025phi000} with the difference that the
directions of the eddy currents and of the secondary magnetic
field are opposite. The same is true for
Fig.~\ref{fig:th100phi000} as compared to
Fig.~\ref{fig:th000phi000}.

\begin{figure}[ht]
    \centering
    \subfigure[]{ 
    \includegraphics[width=0.45\textwidth]{th050phi000.eps}
    \label{fig:th050phi000_1}
    }
    \subfigure[]{ 
    \includegraphics[width=0.45\textwidth]{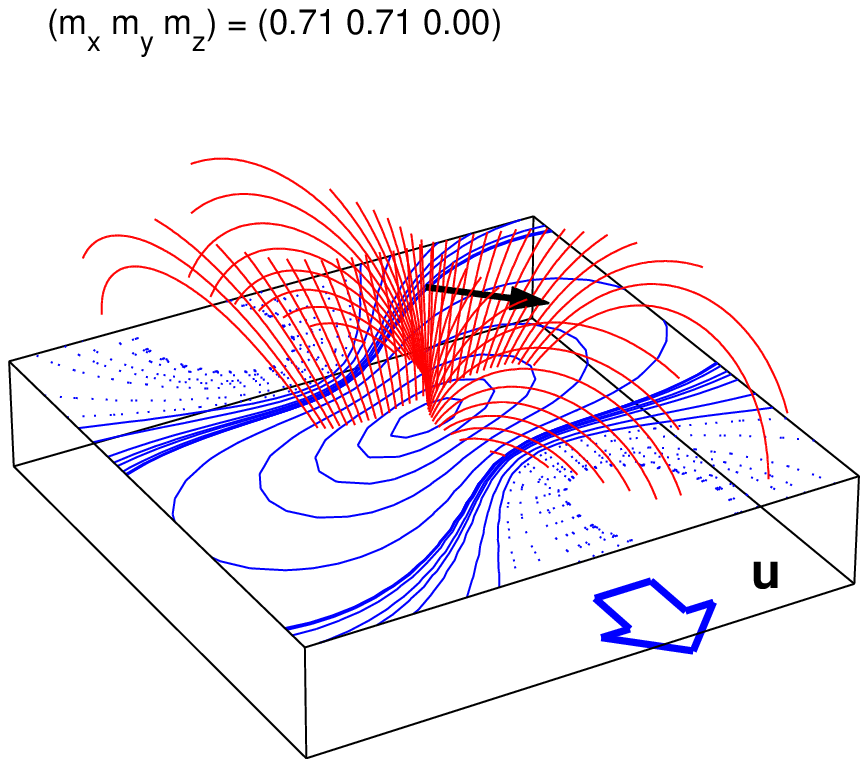}
    \label{fig:th050phi025}
    }\\
    \subfigure[]{
    \includegraphics[width=0.45\textwidth]{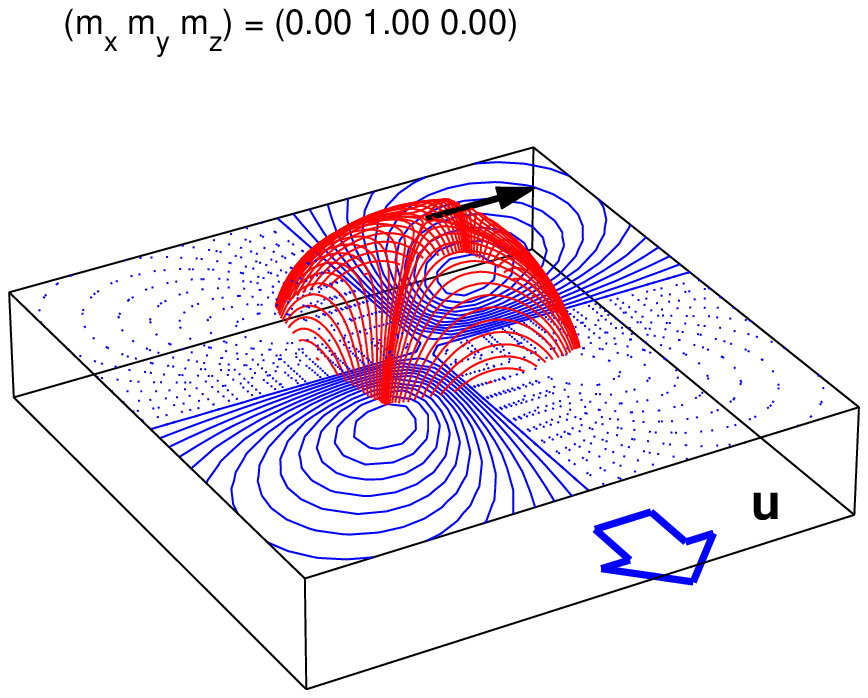}
    \label{fig:th050phi050}
    }\hskip 2.5cm
    \subfigure[]{
    \includegraphics[scale=0.33]{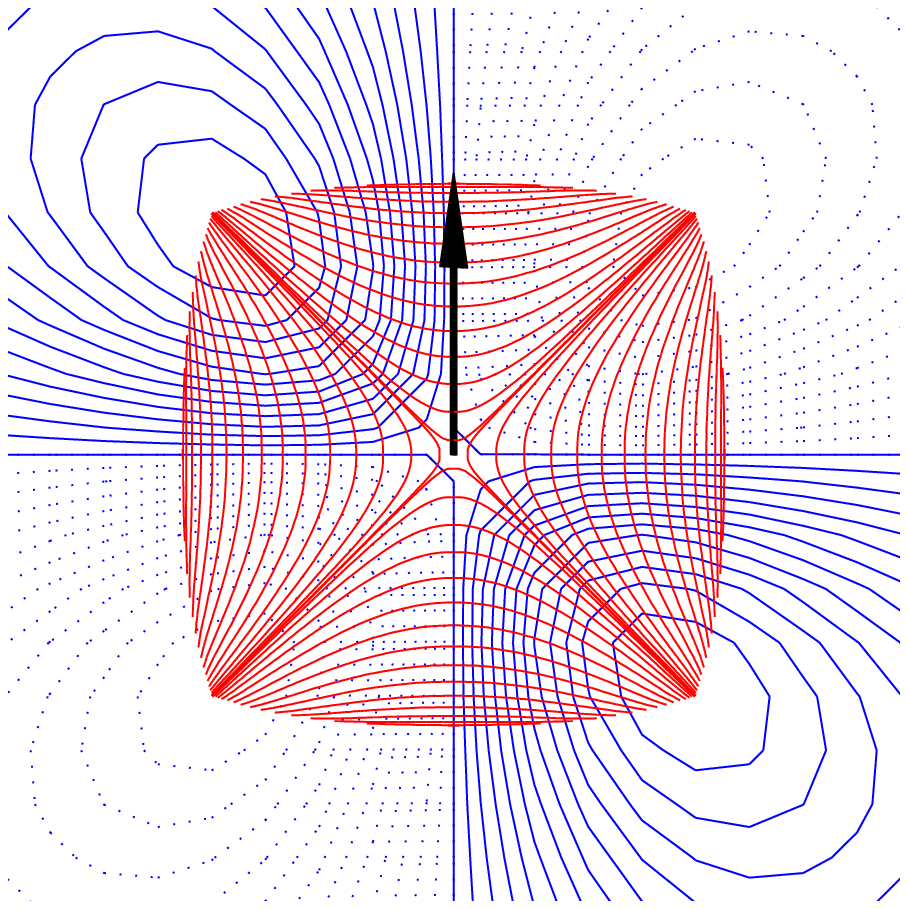}
    \label{fig:th050phi050zoom}
    }\\
    \subfigure[]{
    \includegraphics[width=0.45\textwidth]{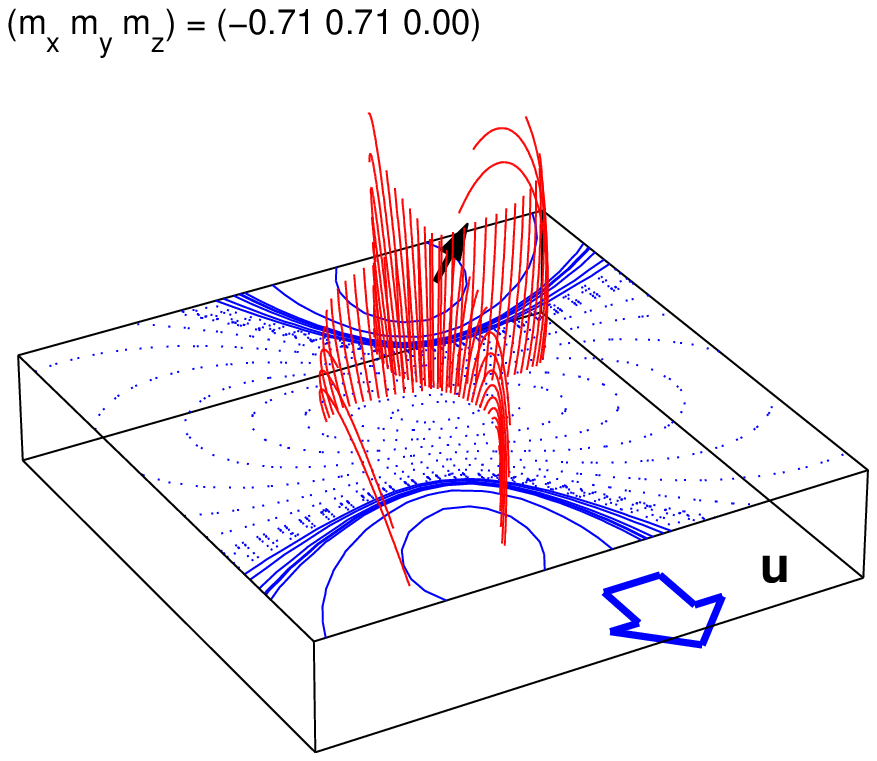}
    \label{fig:th050phi075}
    }
    \subfigure[]{
    \includegraphics[width=0.45\textwidth]{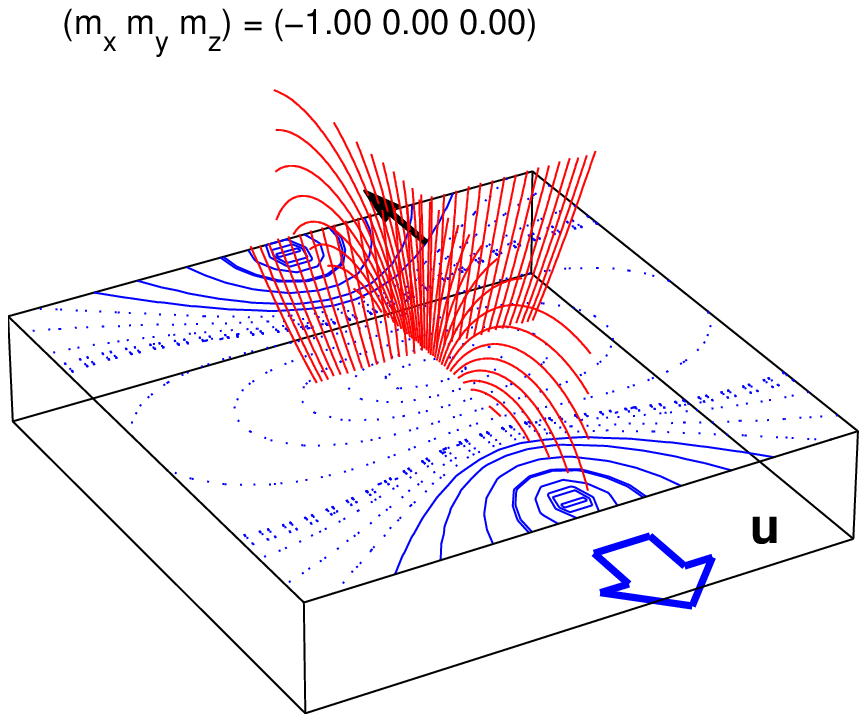}
    \label{fig:th050phi100}
    }
    \caption[Optional caption for list of figures]{
    Eddy currents and secondary magnetic field for the case
    $m_z=0$, i.e. when the magnetic dipole is located in a plane parallel
    to the surface of the moving plate. The points where the secondary
    magnetic field lines emanate are selected manually inside and nearby the
    eddy-current loops  in order to guide the eye.}
    \label{fig:th050}
\end{figure}

Figure~\ref{fig:th050} shows the  behavior at $m_z=0$ and
continuously changing $m_x$ and $m_z$ provided that
$m_x^2+m_y^2+m_z^2=1$. This corresponds to the case when the
magnetic dipole is located in a plane parallel to the surface of
the layer. The case of Fig.~\ref{fig:th050phi000_1}, $m_x=1$,
$m_y=0$, $m_z=0$ is the same as Fig.~\ref{fig:th050phi000} and is
repeated here in order to guide the eye. By rotating the dipole,
the central loop turns  together with the lines of the secondary
magnetic field, Fig.~\ref{fig:th050phi025}. Then, when the main
axis of the central loop is parallel to the diagonal of the plate,
it splits into two equal positive loops, and thus one observes
four alternating eddy currents loops for $m_x=0$, $m_y=1$ $m_z=0$,
see Fig.~\ref{fig:th050phi050}. The magnetic lines emanate from
the positive loop and go into the neighboring negative loop. If
one looks at the system from above, as in
Fig.~\ref{fig:th050phi050zoom} being the same as
Fig.~\ref{fig:th050phi050} but differently oriented, one can see
that the line $r=0$, where the dipole is located, is the place
where the magnetic field lines meet and turn out, holding the
secondary field equal always zero.  Then, the
Fig.~\ref{fig:th050phi075} is a mirror of
Fig.~\ref{fig:th050phi025} and Fig.~\ref{fig:th050phi100} is a
mirror of Fig.~\ref{fig:th050phi000_1} with the difference that
directions of eddy currents and secondary field are opposite in
respect to each other.

\section{Conclusion}\label{sec:conclusion}
An analytic solution is obtained for the electromagnetic
interaction between an arbitrarily oriented magnetic dipole and a
moving electrically conducting solid layer. The solution includes
the electric potential and eddy electric currents inside the layer
as well as the scalar potential for the secondary magnetic field
outside the layer. The formulae obtained are useful for a Lorentz
force velocimetry, because a moving solid plate can be considered
as the limiting case of the mean velocity profile of a turbulent
liquid metal flow when the Reynolds number (Re)  tends to
infinity. It has been shown in \cite{Thess:Votyakov:NJP:2007} that
the signal of a Lorentz force velocimetry system interacting with
a turbulent pipe flow converges towards the signal generated by  a
moving solid cylinder when Re$\rightarrow \infty$.  Hence the
present model provides a benchmark for numerical simulations of
Lorentz force velocimetry in channel flows at high Reynolds
numbers.

It should be finally mentioned that the present model could be
applied in a straightforward way to describe the behavior of an
educational experiment which consists of a spherical permanent
magnet rolling down an inclined aluminium plate. Depending on the
initial orientation of the magnetization the sphere undergoes
different kinds of motion including straight motion with
periodically changing velocity, wiggling motion and irregular
motion. With minor changes the present model could be extended to
cover this case.

\section*{Acknowledgements}

We are grateful to the German Research Foundation (Deutsche
Forschungsgemeinschaft) for supporting EV's visit to Ilmenau in
the framework of the research training group (Graduiertenkolleg)
"Lorentz force velocimetry and Lorentz force eddy current
testing".

\clearpage
\appendix

\section{Eddy currents expressed through a  stream function}
\label{app:stream}
By inserting $\phi(x,y,z)$ from (\ref{eq:potential}) into
(\ref{eq:Ohm}), we find the components of electric current as:
\begin{eqnarray}\label{eq:current}
    j_x &=& \sigma\left\{-\partial_x\phi\right\} = -u\mu\sigma\left\{
    \partial_x\partial_y \partial_n \left[\tanh^{-1}\frac{\Delta z}{R}\right]
    +\partial_x\phi_\infty\right\},
    \\
    j_y &=& \sigma\left\{-\partial_y\phi-u\,B_z\right\} = u\mu\sigma\left\{
    \partial_x\partial_x \partial_n \left[\tanh^{-1}\frac{\Delta z}{R}\right]
    -\partial_y\phi_\infty\right\},
    \\
    j_z &=& 0 ,
\end{eqnarray}
To simplify the above expression for $j_y$, we used $B_z$ from
(\ref{eq:MF:components}) and the fact that
\begin{equation}
    \partial_{xx} \left[\tanh^{-1}\frac{\Delta z}{R}\right] =
    -(\partial_{yy}+\partial_{zz})\left[\tanh^{-1}\frac{\Delta
    z}{R}\right].
\end{equation}The obtained eddy currents are purely horizontal. Due to
their two-dimensionality they can be expressed as
$$
    \mathbf{j}=\nabla\times(\psi\mathbf{e}_z)
$$ where the stream
function $\psi$ is
\begin{eqnarray}\label{eq:psi:intro}
  \psi(x,y,z)&=& -u\mu\sigma\left\{
    \partial_x \partial_n \left[\tanh^{-1}\frac{\Delta
    z}{R}\right]+\psi_\infty(x,y)\right\},\\
    \psi_\infty(x,y) &=& -\lim_{z_{0}\rightarrow\infty}
   \partial_x\partial_n \left[\tanh^{-1}\frac{\Delta
  z}{R}\right]=\sgndelz \,\partial_x\partial_n \ln r.\nonumber
\end{eqnarray}The function $\psi_\infty(x,y)$ is a conjecture of the function $\phi_\infty(x,y)$
and it is needed to nullify eddy currents at the infinite distance
from the dipole.

When the dipole is oriented arbitrarily, the general expression
for $\psi$ can be given explicitly with the aid of the tensor
$\{\bm{\psi}^{(n)}\}=(\psi^{(x)},\psi^{(y)},\psi^{(z)})$ as
follows:
\begin{eqnarray}\label{eq:psi}
    \psi=u\mu\sigma\,\left(\bm{m} \cdot \{\bm{\psi}^{(n)}\}^T\right),
    \hskip 0.5cm
    \{\bm{\psi}^{(n)}\}=
        \left( {\begin{array}{ccc}
            F_0+F_2 f_2 & F_2 g_2  & F_1 f_1
        \end{array} } \right),
\end{eqnarray} where
\begin{eqnarray}\label{eq:fg}
    f_1&=\frac{\delx}{r}, \hskip 1.5cm
    f_k&=f_1 f_{k-1}-g_1 g_{k-1},
    \\
    g_1&=\frac{\dely}{r}, \hskip 1.5cm
    g_k&=g_1 f_{k-1}+f_1 f_{k-1};
\end{eqnarray}and
\begin{eqnarray}\label{eq:Fcoeff}
    F_0=-\frac{\Delta z}{2\,R^3},  \hskip 0.5cm
    F_1=\frac{r}{R^3},\hskip 0.5cm
    F_2=- \frac{\Delta z}{2\,R^3}+\frac{\sgndelz}{r^2}\left[1-\frac{|\Delta
    z|}{R}\right].
\end{eqnarray} The eddy currents $\bm{j}=
u\mu\sigma\,\left(\bm{m}\cdot \{\bm{j}^{(n)}\}^T\right)$ can be
represented with the aid of the tensor $\{\bm{j}^{(n)}\}$ which
is:
\begin{eqnarray}\label{eq:j}
    \{\bm{j}^{(n)}\} =\left( {\begin{array}{ccc}
         C_1 g_1+C_3 g_3 &  C_1 f_1-C_3 f_3 &  C_2 g_2\\
         -3 C_1 f_1-C_3 f_3 &  -C_1 g_1-C_3 g_3 &  C_0-C_2f_2\\
         0 & 0 & 0
         \end{array} } \right),
\end{eqnarray}where
\begin{eqnarray}\label{eq:jcoeff}
    C_0=\frac{3 r^2}{2 R^5}-\frac{1}{R^3}, \hskip 0.5cm
    C_1=\sgndelz \frac{3 r |\delz| }{4 R^5}, \hskip 0.5cm
    C_2=-\frac{3 r^2}{2 R^5},\\
    C_3=\sgndelz \left\{\frac{|\delz|}{r^3}\left[\frac{3 r}{4 R^2}+\frac{1}{r}\right]+
        \frac{2}{r^3}\left[\frac{|\delz|}{R}-1\right]\right\}.
\end{eqnarray}

The coefficients $C_k$ tend to zero when $| \delz |$ goes to
infinity. To find the eddy currents under the dipole, $r=0$, one
has to make a series expansion around  $r=0$. It gives:
\begin{eqnarray} \label{eq:jcoeff01}
 C_0|_{r\rightarrow 0}&= \frac{1}{|\delz|^3}\left\{-1+\frac{3 r^2}{|\delz|^2}\right\}+O\left(r^4\right),
 \\
 C_1|_{r\rightarrow 0}& = \frac{\sgndelz 3 r}{4 |\delz|^4}\left\{1-\frac{5 r^2}{2 |\delz|^2}\right\}+O\left(r^5\right),
 \\
 C_2|_{r\rightarrow 0}& =\frac{3 r^2}{2 |\delz|^5}\left\{-1+\frac{5 r^2}{2 |\delz|^2}\right\}+O\left(r^5\right),
 \\
 C_3|_{r\rightarrow 0}& =\left\{-\frac{\sgndelz\,5 r^3}{8 |\delz|^6}\right\}+O\left(r^5\right)
 \label{eq:jcoeff04}
\end{eqnarray}Thus at $r=0$, $C_0$ is not zero. It is negative, hence, under
the dipole, the direction of the eddy current is opposite to the
$y$-axis, when the dipole is oriented  along the $z$-direction in
agreement with the right-hand rule. These series expansions
results will also be useful afterwards in the analysis of the
secondary magnetic field and forces acting upon the dipole.

It is important to notice that the  functions $f_k$ and $g_k$
above depend on $x$ and $y$ only and do not depend on $z$. This
will be be useful below when we compute secondary magnetic field
by integrating over the plate thickness $z_d\le \hat{z}\le z_u$ .
Moreover, these functions $f_k$ and $g_k$ are specially determined
is such a way, that if the origin of the Cartesian coordinates is
selected on the line passing through the magnetic dipole
perpendicular to the plate, $x_0=y_0=0$, then in the cylindrical
coordinates, $\tan\varphi=y/x$,  $f_k=\cos k\varphi$ and $g_k=\sin
k\varphi$, i.e. the formulae above represent Fourier
decomposition. There must be no angular dependence in a series
expansion around $r=0$ precisely at the point $r=0$, therefore,
$C_0$ is the only non vanishing coefficient in
(\ref{eq:jcoeff01}).

\section{Primitive function for a thin layer}
\label{app:proofA0}
Here, we prove that the solution of (\ref{eq:Ampere:Poisson:A})
for an infinitely thin layer is given by:
\begin{eqnarray}\label{eq:app:A:thinlayer}
  A_0(\hat{z}) &=& \frac{w}{2}\left\{
        \tanh^{-1}\frac{\mhatdelzz{\hatz}}{\hat{R}(\hatz)}+\ln r\right\},
\end{eqnarray} where
\begin{eqnarray}\label{eq:app:A:thinlayer:R}
    \hat{R}(\hatz)^2 &=& r^2+\mhatdelzz{\hatz}^2,
  \hskip 0.3cm \mhatdelzz{\hatz}=|\hatz-z_0|+|\hatz-z|.\nonumber
\end{eqnarray} Function $\mhatdelzz{\hatz}$ can be understood
from the  following geometric interpretation. It is the sum of two
terms, the first term, $|\hatz-z_0|$, is a distance between $z_0$
and $\hatz$, and the second term $|\hatz-z|$ is a distance between
$\hatz$ and $z$. Hence, the layer located at $z=\hatz$ subdivides
the whole space into two half-spaces, $z>\hat{z}$ and $z<\hat{z}$,
therefore the function $\hatdelz(\hat{z})$ in straight brackets
takes the following form:
\begin{eqnarray}\label{eq:dzhat:definition}
    \mhatdelzz{\hatz} =
     \left\{ \begin{array}{ll}
    |z+z_0-2\hat{z}|& \mbox{ if } (\hat{z}-z)(\hat{z}-z_0) > 0;
    \\
    |z-z_0|=|\delz| & \mbox{ if } (\hat{z}-z)(\hat{z}-z_0) < 0.
    \end{array} \right.
\end{eqnarray}The first line is applicable when $z$ and $z_0$ are both in the same
half-space, hence, $\hatdelz(\hatz)=z+z_0-2\hat{z}$ in this case;
 the second line is applicable when $z$ and $z_0$ are in
different half-spaces, hence, $\hatdelz(\hatz)=z-z_0$ in this
case. Then, it is important that $\mhatdelzz{\hatz}$ shows a
discontinuity in the first derivative at $z=\hatz$, i.~e.:
\begin{eqnarray}
    \frac{d}{d z}\,\,\mhatdelzz{\hatz}&=&
     \left\{ \begin{array}{ll}
     +1 & \mbox{if $z >\hat{z}$},\\
     -1 & \mbox{if $z <\hat{z}$},\end{array} \right.
    \hskip 0.3cm \mbox{hence} \hskip 0.3cm
    \frac{1}{2}\,\frac{d^2}{dz^2}\,\, \mhatdelzz{\hatz} =
    \delta(z-\hatz).  \nonumber
\end{eqnarray}Now it is easy to ascertain by means of straightforward calculations
that $\Delta A_0=0$ everywhere except for the point at $z=\hatz$
while at this point the formula (\ref{eq:Ampere:Poisson:A}) is
valid because:
\begin{eqnarray*}\label{eq:A:thin:layer:proof}
    \frac{\partial^2 A_0}{\partial x^2}+\frac{\partial^2 A_0}{\partial y^2} =
    \frac{w}{2}\frac{\mhatdelzz{\hatz}}{\hat{R}^3},
    \\
    \frac{\partial^2 A_0}{\partial  z^2}=\frac{w}{2}\left\{
        -\frac{\mhatdelzz{\hatz}}{\hatR^3}\left[\frac{d}{d z}\,\, \mhatdelzz{\hatz}\right]^2
        + \frac{1}{\hatR}\frac{d^2}{d z^2}\,\,\mhatdelzz{\hatz} \right\},\nonumber\\
    \Delta A_0=\frac{\partial^2 A_0}{\partial x^2}
        +\frac{\partial^2 A_0}{\partial y^2}+\frac{\partial^2 A_0}{\partial z^2}
        =w\,\delta(z-\hatz)\frac{1}{R}, \nonumber
\end{eqnarray*} where the equality $\hatR=R$ at $z=\hatz$ is taken into account.

\section{Tensor for the secondary field of a thin layer}
\label{app:tensor:induced:thin}
As follows from the mapping
(\ref{eq:bxy:thin:layer1}-\ref{eq:bxy:thin:layer3}) between
induced field and eddy currents, and the tensor (\ref{eq:j}) for
the eddy currents, the tensor of the induced field can be written
down as:
\begin{eqnarray}\label{eq:secondaryfield}
  \{\bm{b}^{(n)}\}=\left( {\begin{array}{ccc}
       -3 f_1 P_1-f_3 P_3 &  -g_1 P_1-g_3 P_3 &  f_2 P_2-P_0 \\
      -g_1 P_1-g_3 P_3 & f_3 P_3-f_1 P_1 & g_2 P_2 \\
      P_0-f_2 P_2 & -g_2 P_2 &  -4 f_1 P_1
    \end{array} } \right)
\end{eqnarray} where the coefficients $P_k \equiv w\,P_{0,k}$ for an
infinitely thin layer are obtained from the mapping $P_{0,k}
\leftrightarrow \sgndelzhat C_k/2$, $k=1,\ldots,3$, by keeping in
mind that $\hat{R}$ and $\hat{\Delta}z$ are to be used instead of
$R$ and $\Delta z$. Specifically,
\begin{eqnarray}\label{eq:secondaryfield:coeff:thin}
   P_{0,0} = \sgndelzhat\left\{\frac{3 r^2}{4\hatR^2} -\frac{1}{2
   \hatR^3}\right\},
   \hskip 0.5cm
   P_{0,1} = \frac{3 r \mhatdelz}{8 \hatR^5}, \hskip 0.5cm \\
   P_{0,2} = -\sgndelzhat \frac{3 r^2}{4 \hatR^5}, \hskip 0.5cm
   P_{0,3} =\frac{\mhatdelz}{2\,r^3}\left[\frac{3 r}{4 \hatR^2}+\frac{1}{r}\right]+
        \frac{1}{r^3}\left[\frac{\mhatdelz}{\hatR}-1\right].\nonumber
\end{eqnarray} The series expansion (\ref{eq:jcoeff01})-(\ref{eq:jcoeff04}) is also
applicable for (\ref{eq:secondaryfield:coeff:thin}). One can see
that at $r=0$  $P_{0,0}$  is not vanishing only, therefore, as
follows from (\ref{eq:secondaryfield}), the nonzero tensor
components at $r=0$ are  $b^{(z)}_x= -b^{(x)}_z$ only.

\section{Secondary magnetic field tensor for a half-space
and for a layer with finite thickness}
\label{app:tensor:induced:thick}
For a half-space and for a layer with finite thickness, the
secondary field tensor is of the same structure as for an
infinitely thin layer, Eq.~(\ref{eq:secondaryfield}), with the
following coefficients for the half space:
\begin{eqnarray}\label{eq:Pk:thickplate}
    P_{\infty,0}=-\frac{\hatdelz}{8\hatR^3}, \hskip 1.0cm
    P_{\infty,2}= \sgndelzhat\left\{\frac{\mhatdelz}{8\hatR^3}+\frac{1}{4 r^2}\left[\frac{\mhatdelz}{\hatR}-1\right]\right\},
    \\
    P_{\infty,1}=\frac{r}{16\hatR^3}, \hskip 1.0cm
    P_{\infty,3}= \frac{r}{16\hatR^3}+\frac{1}{4\hatR r} +
       \frac{\sgndelzhat\hatR}{2r^3}\left[\frac{\mhatdelz}{\hatR} -1
       \right],\nonumber
\end{eqnarray} where $\hat{\Delta}z$ and $\hat{R}$ should be computed at
$z_u$.

For the layer with finite thickness, the tensor coefficients are:
\begin{eqnarray}\label{eq:app:Pk:finiteplate}
    P_k(z_u,z_d)= P_{\infty,k}(z_u) - P_{\infty,k}(z_d),
    \hskip 0.5cm k=0\ldots 4;
\end{eqnarray} where $P_{\infty,k}(z_u)$ and
$P_{\infty,k}(z_d)$ are computed according to
(\ref{eq:Pk:thickplate}) with $\hat{\Delta}z$ and $\hat{R}$ taken
at points $z_u$ and $z_d$.

\section{Force field tensor}
\label{app:tensor:force}
The force field tensor, $\{\bm{f}^{(n)}_i\}=\partial_i
\{\bm{b}^{(n)}\}$ can be written down in the following way:
\begin{eqnarray}
    \{\bm{f}^{(n)}_x\} =
      \left( {\begin{array}{ccc}
       3 Q_0+2 f_2 Q_2+f_4 Q_4 &  g_2 Q_2+g_4 Q_4 &  3 f_1 Q_1+f_3 Q_3 \\
       g_2 Q_2+g_4 Q_4 & Q_0-f_4 Q_4 & g_1 Q_1+g_3 Q_3 \\
       -3 f_1 Q_1-f_3 Q_3 & -g_1 Q_1-g_3 Q_3 & 4 Q_0+2 f_2 Q_2
      \end{array} } \right),  \label{eq:forcex} 
      \\
    \{\bm{f}^{(n)}_y\} =
       \left( {\begin{array}{ccc}
        \qquad g_2 Q_2+g_4 Q_4 &  Q_0-f_4 Q_4 &  g_1 Q_1+g_3 Q_3 \\
        \qquad Q_0-f_4 Q_4 &  g_2 Q_2-g_4 Q_4 &  f_1 Q_1-f_3 Q_3 \\
        \qquad -g_1 Q_1-g_3 Q_3 &  f_3 Q_3-f_1 Q_1 &  2 g_2 Q_2
       \end{array} } \right), \label{eq:forcey} 
       \\
    \{\bm{f}^{(n)}_z\} =
       \left( {\begin{array}{ccc}
        \quad -3 f_1 Q_1-f_3 Q_3 &  -g_1 Q_1-g_3 Q_3 &  4 Q_0+2 f_2 Q_2 \\
        \quad -g_1 Q_1-g_3 Q_3 & f_3 Q_3-f_1 Q_1 &  2 g_2 Q_2 \\
        \quad -4 Q_0-2 f_2 Q_2 &  -2 g_2 Q_2 &  -4 f_1 Q_1
       \end{array} } \right). \label{eq:forcez} 
\end{eqnarray}The coefficients $Q_k\equiv w\,Q_{0,k}$ for a infinitely thin layer are:
\begin{eqnarray}\label{eq:Qkforce:thinplate}
    &Q_{0,0}=\frac{3 \mhatdelz}{8 \hat{R}^5}\left\{\frac{5r^2}{2\hatR^2}-1\right\},
    \hskip 0.5cm
    Q_{0,1}=\sgndelzhat\frac{3 r}{2 \hat{R}^5}\left\{\frac{5 r^2}{4\hatR^2}-1\right\},\nonumber
    \\
    &Q_{0,2}=\frac{15 r^2 \mhatdelz}{8 \hat{R}^7},
     \hskip 2.0cm
    Q_{0,3}=\sgndelzhat\frac{15 r^3}{8\hatR^7},
    \\
    &Q_{0,4}=\frac{3\mhatdelz}{2\hatR^3}\left[\frac{5 r^2}{8\hatR^4}
            + \frac{3}{4\hatR^2}+\frac{1}{r^2}\right]
            + \frac{3 }{r^4}\left[\frac{|\hat{\Delta}z|}{\hat{R}}-1
            \right]&.\nonumber
\end{eqnarray}For a half-space, the coefficients  $Q_k\equiv
Q_{\infty,k}$ are obtained by using the same method as done before
for the coefficients $P_{\infty,k}$. This involves an integration
over the thickness of the plate,
$Q_{\infty,k}=\int^{z_u}_{-\infty}Q_{0,k}(\hat{z})\,\,d\hat{z}$,
and  the results of the integration are:
\begin{eqnarray}\label{eq:Qkforce:thickplate}
    Q_{\infty,0}=\frac{1}{16\hatR^3}\left\{\frac{3 r^2}{2\hatR^2}-1\right\},
    \hskip 0.75cm\\
    Q_{\infty,1}=-\sgndelzhat\frac{3 r \mhatdelz}{16 \hatR^5},
    \hskip 0.75cm
    Q_{\infty,2}=\frac{3 r^2}{16\hatR^5},
    \\
    Q_{\infty,3}=-\sgndelzhat\left\{\frac{\mhatdelz}{4r\hatR^3}
        \left[1+\frac{3 r^2}{4\hatR^2}\right]
        +\frac{1}{2r^3}\left[\frac{\mhatdelz}{\hatR}-1\right]\right\},
    \\
    Q_{\infty,4}=\frac{3}{4 \hatR r^2}
    \left[1+\frac{r^2}{4\hatR^2}+\frac{r^4}{8\hatR^4}\right]
        +\frac{3 \hatR}{2r^4}\left[\frac{\mhatdelz}{\hatR}-1\right].
\end{eqnarray}For a plate with finite thickness we have:
$$
    Q_{k}=Q_{\infty,k}(z_u) - Q_{\infty,k}(z_d), \hskip 0.5cm k=0\ldots 4.
$$

\section{Non-zero tensor coefficients at $r=0$ and $z=z_0$} \label{app:P0Q0}
The coefficients for a thin layer of thickness $w\ll|z_0-\hatz|$
located at $\hatz$ are:
\begin{eqnarray}
    \left.P_0\right|_{r=0,z=z_0}\equiv w\left.P_{0,0}\right|_{r=0,z=z_0}
        &=-w\frac{\mbox{sgn}(z_0-\hatz)}{16|z_0-\hatz|^3},
        \label{eq:P0:thin}
    \\
    \left.Q_0\right|_{r=0,z=z_0}\equiv w\left.Q_{0,0}\right|_{r=0,z=z_0}
        &=-w\frac{3}{128\,|z_0-\hatz|^4}. \label{eq:Q0:thin}
\end{eqnarray} The same coefficients for a half-space are
obtained by integration over $\hatz$ from $z_u$ up to infinity
\begin{eqnarray}
    \left.P_0\right|_{r=0,z=z_0}\equiv \left.P_{\infty,0}\right|_{r=0,z=z_0}
        &=-\frac{1}{32\,|z_0-z_u|^2}, \label{eq:P0:thick}
    \\
    \left.Q_0\right|_{r=0,z=z_0}\equiv \left.Q_{\infty,0}\right|_{r=0,z=z_0}
        &=-\frac{1}{128\,|z_0-z_u|^3}. \label{eq:Q0:thick}
\end{eqnarray} Finally, for a plate with finite thickness, $z_d \le \hatz
\le z_u$:
\begin{eqnarray}
    \left.P_0\right|_{r=0,z=z_0} &=\frac{1}{32}
        \left\{\frac{1}{|z_0-z_d|^2}-\frac{1}{|z_0-z_u|^2}\right\},\label{eq:P0:finite}
    \\
    \left.Q_0\right|_{r=0,z=z_0} &=\frac{1}{128}
        \left\{\frac{1}{|z_0-z_d|^3}-\frac{1}{|z_0-z_u|^3}\right\}.\label{eq:Q0:finite}
\end{eqnarray}

\section*{References}
\bibliographystyle{plain}
\bibliography{./../../Bibtex/mhd}

\end{document}